\documentclass[
showpacs,
 amsmath,amssymb,
 aps,
twocolumn,
eqsecnum,
]{revtex4}
\usepackage{epsfig}
\usepackage{euscript,graphicx}
\usepackage{amsthm,dsfont,amsfonts,amsmath,amssymb}
\usepackage{euscript,color,fontenc,textcomp,relsize}
\usepackage{bm,url,float,cleveref}
\usepackage[caption=false]{subfig}
\usepackage[english]{babel}
\usepackage{natbib}
\allowdisplaybreaks

\begin{document}
\title{Analytic expression of perturbations of Schwarzschild spacetime\\via Homotopy Analysis Method}
\date{\today}
\author{Gihyuk Cho$^{1}$\footnote{croakerhyuk@gmail.com}}
\affiliation{$^{1}$
Korea Astronomy and Space Science Institute,
776 Daedeokdaero, Daejon, Korea
 }

\begin{abstract}
This paper derives the analytic and practicable expression of general solution of vacuum Regge-Wheeler equation via Homotopy Analysis Method.
\end{abstract}
\maketitle
\section{Introduction}
Since the blackhole perturbation theory was given its birth from the investigation on the stability of Schwarzschild metric by Regge $\&$ Wheeler~\cite{Regge1957} and Vishveshwara~\cite{Vishveshwara1970}, there have been much development (See \cite{Chandrasekhar1985} for a comprehensive review). The development for perturbed Schwarzchild metric could be summarized into the two pieces: (1) the 6 gauge-invariant description~\cite{Moncrief1974,Gerlach1980,Thompson2017} and (2) the (generalized) Darboux transformation~\cite{Chandrasekhar1975, Glampedakis2017}. The first piece states that 10 components of the metric perturbation compose 6 independent variables which are \textit{gauge invariant} up to linear order gauge transformation, and the resulting 6 coupled equations governing these 6 gauge-invariants, are reduced down to two master equations called Regge-Wheeler and Zerilli equations \cite{Regge1957,Zerilli1970}.(Regge-Wheeler and Zerilli functions are also gauge invariant.) Although, in the original papers of Regge $\&$ Wheeler and Zerilli, they chose specific gauges (called Regge-Wheeler and Zerilli gauges) to derive the equations, one can now construct Regge-Wheeler and Zerilli equations in any gauges. Given that 4 more equations regarding a gauge condition are of presence, one can reconstruct 10 components of the perturbed metric. Secondly, as long as vacuum (or, homogeneous) solutions are of interest, Regge-Wheeler and Zerilli functions are connected by Chandraseckhar transformation~\cite{Chandrasekhar1975} which is one of Darboux transformations~\cite{ Glampedakis2017}. Practically, solving non-homogeneous Regge-Wheeler/Zerilli equations could be easily done, once homogeneous solutions are known, by means of Green's function method. From this understanding, we came up with an idea that getting homogeneous Regge-Wheeler/Zerilli equation solved into a closed form expression might be crucial to further development of the blackhole perturbation theory. (What is explained in this paragraph is formulated in Appendix.\ref{app:Rec}).\\
\indent We are not the first one who tried to solve them into analytic expressions. In 1996, Shuhei Mano et al. solved Teukolsky equation by two series of special functions, Hypergeometric/Coulumb wave functions~\cite{Mano1996,Sasaki2003}. They are really exact solutions but too complicated and slowly converging to be used for practical purposes. On the other hand, there is another approach of making use of confluent Heun function~\cite{Fiziev2006}. However, we do not see confluent Heun function as an analytic expression because any integral representation and series representation convergent at arbitary point, are yet to be known~\cite{Motygin2018}. Here, we are going to derive fast and arbitrarily converging expression for Regge-Wheeler function, and it is possibly done by Homotopy Analysis Method (HAM)~\cite{Liao2003,Liao2011}. HAM is simple in priniciple, but pratically it has too many options to pose. Hence it is not trival to find a good choice so that not only converging resulting series is, but also it converges fast enough to be practicable in use. In the case of blackhole perturbation theory, only after many trials have been conducted and failed, we managed to discover the setting where HAM works practicably. For examples, we made several attempts to solve Zerilli's equation and Teukolsky's equation, but failed. When it comes to Regge-Wheeler equation, HAM finally works after a change of variable is performed. This is why we choose Regge-Wheeler function for a ground.\\
\indent In this paper, we present and derive the analytic solution of Regge-Wheeler equation up to the third HAM order. Before solving Regge-Wheeler equation, we are going to provide a brief review on how HAM works in \S.\ref{sec:HAM}. And,  \S.\ref{sec:RW} poses the essential formalism of this paper, such as the deformation equation and the initial guess. In \S.\ref{sec:First}, \S.\ref{sec:Second} and \S.\ref{sec:Third}, we perform the first, second and third order integrations. In \S.\ref{sec:hbar}, we determine numerical value of $\hbar$ with an empirical procedure which will be explained, and make a comparison with numerical integrations and discuss its limits.
\section{Homotopy Analysis Method}\label{sec:HAM}
This section provides a brief review on Homotopy Analysis Method (HAM) (See \cite{Liao2003,Liao2011} for a detailed review on HAM). In spite of the fact that HAM also works for partial differential equation of an arbitrary order, for brevity, let us suppose that one needs to solve an ordinary and second-order differential equation,
\begin{align}
\mathcal{E}[u(x)]=0\,,
\end{align}
where $\mathcal{E}$ represents a differential operator, hard to solve. With a certain boundary condition $\mathcal{B}$, let $u_0(x)$ be an initial guess of the exact solution $u(x)$, and satisfy $\mathcal{B}$. After choosing a linear second-order differential operator $\mathcal{L}$, which you are confident in solving, one can write down the following what is called \textit{deformation equations},
\begin{align}\label{HAMeq}
(1-q)\,\mathcal{L}\big[U(x;q)-u_0(x)\big]=q\,\hbar\,\mathcal{E}\big[U(x;q)\big]\,,
\end{align}
where $q\in[0,1]$ is a real number, called deformation parameter, and $\hbar$ is a non-zero complex/real number. Let $U(x;q)$ be the exact solution of Eq.(\ref{HAMeq}). It is easily found that $U(x;0)$ and $U(x;1)$ satisfy
\begin{subequations}
\begin{align}
\mathcal{L}&\big[U(x;0)-u_0(x)\big]=0\,,\\[1ex]
\mathcal{E}&\big[U(x;1)\big]=0\,,
\end{align}
\end{subequations}
respectively. Since $u_0$ is assumed to satisfy $\mathcal{B}$ already, it is followed that $U(x;0)=u_0$, while $U(x;1)=u(x)$ (the exact solution). Thus, $U(x;q)$ could be seen as a device deforming the initial guess $u_0(x)$ to the exact one $u(x)$ as $q$ runs from $0$ to $1$.\\
\indent Instead of solving Eq.(\ref{HAMeq}) directly, we solve it in Taylor expanded form about $q=0$, order by order,
\begin{align}\label{HAMsol}
U(x;q)=u_0+\sum^\infty_{n=1}\,q^n\,u_n(x)\,.
\end{align}
For instance, the equation for $u_1$ is
\begin{align}
\mathcal{L}[u_1]=\hbar\,\mathcal{E}[u_0]\,.
\end{align}
After all computations, explicit value of  $\hbar$ is adjusted to accelerate the convergence. Note that it is proved in \cite{Liao2003} (See \textit{Theorem 2.1.}  therein) that if Eq.(\ref{HAMsol}) is convergent then it converges to the exact solution.\\

\section{Formalism and Nomenclature}\label{sec:RW}

\subsection{Deformation equation}
Regge-Wheeler function $Q_{l\omega}(r)$ is a solution of the Regge-Wheeler differential equation in the Schwarzchild coordinate,
\begin{align}\label{RWeq}
\frac{d^2Q_{l\omega}(r)}{dr^{*2}}+\bigg[\,\omega^2-\Big(1-\frac{2}{r}\Big)\,\Big(\frac{l\,(l+1)}{r^2}-\frac{6}{\,r^3}\Big)\bigg]\,Q_{l\omega}(r)=0\,,
\end{align}
where $r^*=r+2\,\log(r-2)$. When $r\to\infty (r^*\to\infty)$, the equation becomes the ordinary wave equation,
\begin{align}
\frac{d^2}{dr^{*2}}Q_{l\omega}(r)+\omega^2\,Q_{l\omega}(r)=0\,,
\end{align}
of which two independent solutions should be $\text{e}^{i\,r^*\,\omega}$ and $\text{e}^{-i\,r^*\,\omega}$. Instead of considering both cases, we only consider the \textit{outgoing-wave} boundary condition,
\begin{align}\label{bd}
\lim_{r\to\infty}\,Q_{l\omega}(r)=\text{e}^{i\,r^*\,\omega}\,.
\end{align}
Hereafter, we will denote the outgoing-wave solution just as $Q(r)$ for convenience, without $\omega$, $l$ symbols and mentioning its boundary condition Eq.(\ref{bd}) anymore.\\
For the linear operator $\mathcal{L}$, we make a simplest choice as
\begin{align}
\mathcal{L}=\frac{d^2}{dr^2}\,.
\end{align}
However, solving for $Q$ itself was not effective. Instead, we introduce \begin{align}
\mathcal{F}:=Q\,\text{e}^{-2\,I\,\omega\,\log(r-2)}\,,
\end{align}
which leads to the deformation equation,
\begin{align}
(1-q)\,\frac{d^2}{dr^2}\Big[\,\mathcal{F}(r;q)-\mathcal{F}_0(r)\Big]=q\,\hbar\,\mathcal{E}\big[\mathcal{F}(r;q)\big]\,,
\end{align}
where
\begin{align}
\mathcal{E}&=\frac{(r-2)}{r}\frac{d^2}{dr^2}+\frac{(2+4\, i\, r\, \omega )}{r^2}\,\frac{d}{dr}\,\\[0.5ex]\notag
&+\frac{\left(-l^2\, r-l\, r+r^3 \omega ^2+2 \,r^2\, \omega ^2-2\, i\, r\, \omega +6\right)}{r^3} \,.
\end{align}
Expanding it about $q=0$ with $\mathcal{F}(r;q)=\sum_{n=0}^\infty \,q^n\, \mathcal{F}_n(r)$ yields the following subequations up to $q^3$ order,
\begin{subequations}
\begin{align}\label{eqs}
\frac{d^2}{dr^2}\mathcal{F}_1(r)&=\hbar\,\mathcal{E}[\mathcal{F}_0]\,,\\[1ex]
\frac{d^2}{dr^2}\mathcal{F}_2(r)&=\frac{d^2}{dr^2}\mathcal{F}_1+\hbar\,\mathcal{E}[\mathcal{F}_1]\,,\\[1ex]
\frac{d^2}{dr^2}\mathcal{F}_3(r)&=\frac{d^2}{dr^2}\mathcal{F}_2+\hbar\,\mathcal{E}[\mathcal{F}_2]\,.
\end{align}
\end{subequations}
For convenience, we introduce the following parametrization,
\begin{subequations}
\begin{align}
\mathcal{G}_n&:=\mathcal{E}[\mathcal{F}_n]\,,\\[1ex]
\mathcal{H}_{n+1}&:=\int^r_\infty dr'\,\mathcal{G}_n(r')\,,\\[1ex]
\mathcal{I}_{n+1}&:=\int^r_\infty dr'\,r'\,\mathcal{G}_n(r')\,.
\end{align}
\end{subequations}
This parametrization has an advantage that the derivatives are simply expressed,
\begin{subequations}
\begin{align}
\frac{d}{dr}\mathcal{F}_1&=\hbar\,\mathcal{H}_1\,,\\[1ex]
\frac{d}{dr}\mathcal{F}_2&=\hbar\,\big(\mathcal{H}_1+\mathcal{H}_{2}\big)\,,\\[1ex]
\frac{d}{dr}\mathcal{F}_3&=\hbar\,\big(\mathcal{H}_1+\mathcal{H}_{2}+\mathcal{H}_{3}\big)\,.
\end{align}
\end{subequations}
\subsection{Initial guess: Superasymptotic Solution}
In order to get a HAM solution of Eq.(\ref{RWeq}), we need an explicit expression of the initial guess $\mathcal{F}_0$ to start with. 
\begin{align}
\mathcal{F}_0(r)&:=\text{e}^{i\,\omega\,r}\sum_{n=0}^M \frac{b_n}{r^n}\,,
\end{align}
where $M\geq0$ is a non-negative integer not a mass. Now that we have chosen a series solution, it is needed to get the recurrence relation for $b_n$ as
\begin{align}\label{rec}
b_n&= \frac{i\, \big[\,(l-1)\,(l+2)-(n+1)\,(n-2)\,\big] \,b_{n-1}}{2\,n\,\omega }\notag\\[1ex]
&+\frac{i\, (n+1)\,(n-3)\,b_{n-2}}{ n\, \omega}\,,
\end{align}
with $b_{-2}=b_{-1}=0$ and $b_0=1$ by which the boundary condition Eq.(\ref{bd}) holds. Now that this recurrence relation reveals that $b_n\sim\frac{i n}{\omega}$ diverges as $n\to\infty$, we can conclude that the asymptotic solution is not convergent. However, this still provides a good approximation when it is truncated at an appropriate order as many other asmyptotic series often do. Our numerical investigation empirically reveals that when $M=l+2,  l+3$, the asymptotic solution gives the best agreement. And also there is another tendency that the larger $\omega$ is, the larger $M$ is required. However, it is not necessary, so we will keep $M$ as a variable with a constraint $M\geq4$.
\section{First Order Integration}\label{sec:First}
In this subsection, we are going to get $\mathcal{F}_1(r)$. Since $\mathcal{F}_0$ already satisfies the boundary condition we are imposing that $\mathcal{F}_1$ and its derivative both must be zero at infinity so that $\mathcal{F}_1$  does not ruin the overall boudary condition.
\begin{subequations}\label{bdF1}
\begin{align}
\mathcal{F}_1\big{|}_{r=\infty}&=0\,,\\[1ex]
\frac{d\mathcal{F}_1}{dr}\bigg{|}_{r=\infty}&=0\,.
\end{align}
\end{subequations}
Using the recursion Eq.(\ref{rec}), it is easily derived that
\begin{widetext}
\begin{subequations}
\begin{align}
\mathcal{G}_0&=\mathcal{A}\,\frac{\text{e}^{i\, r \,\omega }}{r^{M+2}}+\mathcal{B}\,\frac{\text{e}^{i\, r \,\omega }}{r^{M+3}}\,,
\end{align}
where
\begin{align}
\mathcal{A}&=b_M\, \big(M+l+1\big)\,\big(M-l\big)-2 \,b_{M-1}\,\big(M+2\big)\,\big(M-2\big)\,,\\[1ex]
\mathcal{B}&=-2  \,b_M\,\big(M-1\big)\,\big(M+3\big)\,.
\end{align}
\end{subequations}
Indentfying some integrals with incomplete Gamma function,
\begin{align}
\Gamma(-n,z):=\int^\infty_z\,t^{-n-1}\,\text{e}^{-t}\,dt\,,\quad\text{($n$ is a non-negative integer.)}
\end{align}
the result is derived as below:
\begin{subequations}\label{HI1}
\begin{align}\label{F1}
\mathcal{F}_1(r)&=\hbar\,\big(r\,\mathcal{H}_1-\mathcal{I}_1\big)\,,
\end{align}
where
\begin{align}
\mathcal{H}_1&=(-i)^M\,\omega^{M+1}\,\bigg(i\,\mathcal{A}\,\,\Gamma(-M-1,-i\,r\,\omega)+\omega\,\mathcal{B}\,\,\Gamma(-M-2,-i\,r\,\omega)\bigg)\,,\\[2ex]
\mathcal{I}_1&=(-i)^{M-1}\,\omega^{M}\,\bigg(i\,\mathcal{A}\,\,\Gamma(-M,-i\,r\,\omega)+\omega\,\mathcal{B}\,\,\Gamma(-M-1,-i\,r\,\omega)\bigg)\,.
\end{align}
\end{subequations}\\
Apparently, it seems that the incomplete Gamma function is not any closed form expression. However, by the following series representation, which converges in entire complex plane $|z|<\infty$, we can call $\Gamma(n,z)$ analytic  \cite{Gammaseries},
\begin{align}\label{igamma}
\Gamma(-n, z)=\frac{(-1)^n}{n!}\Big(\psi(n+1)-\log z\Big)-z^{-n}\sum_{k=0, k\neq n}^\infty\frac{(-z)^k}{(k-n)\,k!}\,,
\end{align}
where $\psi(z)=\frac{\Gamma'(z)}{\Gamma(z)}$ is the polygamma function. As $|z|$ increases, more terms are required for a certain accuracy. Since one needs infinite number of terms to see its asymptotic behaviour, we just write down its asmyptotic behaviour \cite{Gammaasymp} at $|z|\to\infty$,
\begin{align}\label{asymGamma}
\Gamma(-n, z)=\text{e}^{-z}\,z^{-n-1}\,\Big(1+\mathcal{O}(1/z)\Big)\,.
\end{align}
With this asymptotic behaviour, we can conclude that the above expression of $\mathcal{F}_1$, Eq.(\ref{F1}), satisfies the boundary conditions Eqs.(\ref{bdF1}).
\section{Second order Integration}\label{sec:Second}
Now, we are in the position of solving the second order equation. First, we get
\begin{subequations}\label{HI2}
\begin{align}
\mathcal{G}_1=\text{e}^{i\, r \,\omega } \,\bigg(\,\frac{\text{$\gamma_0 $}}{r^M}+\frac{\text{$\gamma_1 $}}{r^{M+1}}+\frac{\text{$\gamma _2$}}{r^{M+2}}+\frac{\text{$\gamma_3 $}}{r^{M+3}}+\frac{\text{$\gamma_4 $}}{r^{M+4}}\,\bigg)+\Gamma (-M,-i \,r \,\omega ) \left(\text{$\delta_0 $}\, r+\text{$\delta_1 $}+\frac{\text{$\delta_2 $}}{r}+\frac{\text{$\delta_3 $}}{r^2}+\frac{\text{$\delta_4$}}{r^3}\,\right)\,,
\end{align}
where the coefficients are displayed in Appendix.~\ref{app:Coe}. Again, with the following boundary conditions,
\begin{align}
\mathcal{F}_2\big{|}_{r=\infty}&=0\,,\\[1ex]
\frac{d\mathcal{F}_2}{dr}\bigg{|}_{r=\infty}&=0\,,
\end{align}
\end{subequations}
we perform the integration for Eq.(\ref{eqs}). The result is
\begin{subequations}
\begin{align}
\mathcal{F}_2=\mathcal{F}_1+\hbar\,\Big(\,r\,\mathcal{H}_2-\mathcal{I}_2\,\Big)\,,
\end{align}
\begin{align}
\mathcal{H}_{2}(r)&=\delta_2\,U(r)+\text{e}^{i\,r\,\omega}\,\bigg(\frac{\epsilon_0}{r^{M-1}}+\frac{\epsilon_1}{r^{M}}+\frac{\epsilon_2}{r^{M+1}}+\frac{\epsilon_3}{r^{M+2}}+\frac{\epsilon_4}{r^{M+3}}\bigg)\\[1ex]\notag
&+\Gamma(-M,-i\,r\,\omega)\bigg(\text{$\zeta_0 $} \,r^2+\text{$\zeta_1$}\, r+\zeta_2+\frac{\text{$\zeta_3 $}}{r}+\frac{\text{$\zeta_4$}}{r^2}\bigg)\,,\\[2ex]
\mathcal{I}_{2}(r)&=\delta_3\,U(r)+\text{e}^{i\,r\,\omega}\,\bigg(\frac{\eta_0}{r^{M-2}}+\frac{\eta_1}{r^{M-1}}+\frac{\eta_2}{r^{M}}+\frac{\eta_3}{r^{M+1}}+\frac{\eta_4}{r^{M+2}}\bigg)\\[1ex]\notag
&+\Gamma(-M,-i\,r\,\omega)\,\bigg(\text{$\theta_0 $} \,r^3+\text{$\theta_1$}\, r^2+\theta_2\,r+\theta_3+\frac{\text{$\theta_4 $}}{r}\bigg)\,,
\end{align}
\end{subequations}
where the explicit expressions of every coefficients can be also found in Appendix.\ref{app:Coe}. \\
In order to clarify the process of the integration, we display the elementary integrals used here:
\begin{subequations}\label{integralrules}
\begin{align}
\int^{r}_{\infty}{\text{e}^{i\,r'\,\omega}}{r'^n}\,dr'&=\frac{\Gamma (n+1,-i \,r \,\omega)}{
(i \,\omega )^{n+1}} \quad(\text{if $n<-1$})\,,
\end{align}
\begin{align}
\int^{r}_\infty \Gamma(-M,-i\,r'\,\omega)\,r'^n\,dr'&=\frac{r^{n+1}\, \Gamma (-M,-i\,r\,\omega )+(-i\,\omega)^{-n+1}\,\Gamma (-M+n+1,-i\,r\,\omega )}{n+1} \quad (\text{if}\, n\neq-1\,\&\&\,M\geq n+1)\,.
\end{align}
\end{subequations}
However, the integration $\int \frac{\Gamma(-M,-i\,r\,\omega)}{r}\,dr$ is not trivial. We use the following identity with a non-negative integer $N$,
\begin{align}
\Gamma(-N,z)=\frac{1}{N!}\bigg[\frac{\text{e}^{-z}}{z^N}\,\sum_{k=0}^{N-1}(-1)^k(N-k-1)!\,z^k+(-1)^N\,\Gamma(0,z)\bigg]\,.
\end{align}
As invoking the following integration, 
\begin{align}
\int dz\frac{\Gamma(0,z)}{z}=z \,_3F_3(1,1,1;2,2,2;-z)-\frac{1}{2} \log z\, (\log z+2 \,\gamma_E )\,,
\end{align}
where $\gamma_E=0.577216\cdots$ is the \textit{Euler–Mascheroni} constant, and $_3F_3(1,1,1;2,2,2;z)$ is the generalized hypergeometric function of which series representation \cite{HPQseries} is given as
\begin{align}\label{33series}
_3F_3(1,1,1;2,2,2;z)=\sum_{k=0}^{\infty}\frac{z^k}{(k+1)^3\,k!}\,,
\end{align}
we define a new funcniton $\mathcal{U}(r)$ as
\begin{align}\label{calU}
\mathcal{U}(r):=-i\,r\,\omega\,_3F_3(1,1,1;2,2,2;i\,r\,\omega)-\frac{1}{2}\,\log(-i\,r\,\omega)\,\big(\log(-i\,r\,\omega)+2\,\gamma_E\big)-\frac{1}{2}\,\Big(\gamma_E^{\,2}+\frac{\,\pi^2}{6}\Big)
\end{align}
Since the series Eq.~(\ref{33series}) converges for all complex number $z$, ($|z|<\infty$) and very fast, we use the series representation for practical comptutations. However, series representation is not suitable for looking for asymptotic behaviour. According to Appendix.\ref{app:AsympU}, where its asymptote is investigated from the integral representation, the last numerical term $\frac{1}{2}\,\Big(\gamma_E^{\,2}+\frac{\,\pi^2}{6}\Big)$ in Eq.(\ref{calU}) has been introduced so that
\begin{align}
\lim_{r\to\infty}\mathcal{U}(r)=0\,.
\end{align}
Now we can integrate $\frac{\Gamma(-M,-i\,r\,\omega)}{r}$ as
\begin{align}
U(r)&:=\int^{r}_{\infty} \,\frac{\Gamma(-M,-i\,r\,\omega)}{r}\,dr\,,\\[1ex]\notag
&=\sum_{k=0}^{M-1}\frac{\,(-1)^{k+1}}{M!}(M-k-1)!\,\Gamma(k-M,-i\,r\,\omega)+\frac{\,(-1)^M}{M!}\,\mathcal{U}(r)\,,\notag\\[1ex]
&=\frac{\text{e}^{i\, r \,\omega }}{M!} \sum_{p=1}^{M-1}\bigg(\frac{(-1)^{p+1}\,(M-p)!}{(-i\,r\,\omega)^{M+1-p}}  \,\psi_{M-p+1}\bigg)-\psi_{M+1}\,\Gamma (-M,-i \,r\, \omega )+\frac{\,(-1)^M}{M!}\mathcal{U}(r)\,.\notag
\end{align}
Here, $\psi_N$ ($\psi$ with a subscript) is not the polygamma function $\psi(N)$, but
\begin{align}
\psi_N:=\sum_{k=0}^{N-1}\frac{1}{k}=\psi(N)+\gamma_E\,.
\end{align}
\section{Third and higher order Integration}\label{sec:Third}
From the third order integration, the integrands are not bounded at infinity ($r=\infty$), so the intergrations are not bounded as well. However, one can extend this formalism up to some higher orders discarding unbounded region. As an example, this section derives the HAM solution of the third order without specifying boudary conditions and suggest how to avoid the non-finite integrations. Because of the linearity of the system of our interest, $\mathcal{G}_2$ includes what we already derived,
\begin{align}
\mathcal{G}_2&=\mathcal{E}[\mathcal{F}_1+\hbar\,\big(r\,\mathcal{H}_2-\mathcal{I}_2\big)]=\mathcal{G}_1+\hbar\,\mathcal{E}[r\,\mathcal{H}_2-\mathcal{I}_2]\,.
\end{align}
Let us denote the second term as $\hat{\mathcal{G}}_{2}:=\mathcal{E}[r\,\mathcal{H}_2-\mathcal{I}_2]$. Consecutively, it is followed that
\begin{subequations}
\begin{align}
\mathcal{F}_3
&=\mathcal{F}_2+\hbar\,(r\,\mathcal{H}_3-\mathcal{I}_3)\,,\notag\,\\[1ex]
&=-\mathcal{F}_1+2\,\mathcal{F}_2+\hbar^2\,(r\,\hat{\mathcal{H}}_3-\hat{\mathcal{I}}_3)\,,
\end{align}
where the parameters are given as
\begin{align}
\mathcal{H}_3&=\mathcal{H}_2+\hbar\,\underbrace{\int^r\,dr'\,\hat{\mathcal{G}_2}(r')}_{=:\,\hat{\mathcal{H}}_3}\,,\\[1ex]
\mathcal{I}_3&=\mathcal{I}_2+\hbar\,\underbrace{\int^r\,dr'\,\hat{r'\,\mathcal{G}_2}(r')}_{=:\,\hat{\mathcal{I}}_3}\,.
\end{align}
\end{subequations}
As same as previous, we list the results of integrations below:
\begin{subequations}
\begin{align}
\hat{\mathcal{H}}_3(r)&=\text{e}^{i\,r\,\omega}\,\sum_{i=0}^{6}\frac{\kappa_i}{r^{M-2+i}}+\Gamma(-M,-i\,r\,\omega)\,\sum_{i=0}^{6}\frac{\lambda_i}{r^{-4+i}}+\Omega_{3HU}\,U(r)+\Omega_{3HV}\,V(r)\,,\\[1ex]
\hat{\mathcal{I}}_3(r)&=\text{e}^{i\,r\,\omega}\,\sum_{i=0}^{6}\frac{\xi_i}{r^{M-3+i}}+\Gamma(-M,-i\,r\,\omega)\,\sum_{i=0}^{6}\frac{\rho_i}{r^{-5+i}}+\Omega_{3IU}\,U(r)+\Omega_{3IV}\,V(r)\,.
\end{align}
\end{subequations}
The explict expressions for the coefficients which appear here, are too lengthy to write down, we list them in the enclosed \textsc{Mathematica} file in the name of $H3hat$ and $I3hat$. The additional elementary integrals used for the third order integration are as below:
\begin{align}
\int^{r}\,dr'\,r'^{n}\,U(r')=\frac{U(r)\,r^{n+1}}{n+1}-\frac{r^{n+1}\, \Gamma (-M,-i\,r\,\omega )+(-i\,\omega)^{-n+1}\,\Gamma (-M+n+1,-i\,r\,\omega )}{(n+1)^2}\,,\\ \quad (\text{if}\, n\neq-1\,\&\&\,M\geq n+1)\,.\notag
\end{align}
The most tricky part is
\begin{align}
&V(r):=\int^{r}\,dr'\,\frac{U(r')}{r'}\,,\\[2ex]
&=\int^{r}\,dr'\,\Bigg[ \sum_{p=1}^{M-1}\bigg(\frac{(-1)^{p+1}\,(M-p)!\,\psi_{M-p+1}\,\text{e}^{i\, r \,\omega }}{(-i\,\omega)^{M+1-p}\,M!\,r^{M+2-p}}  \,\bigg)-\psi_{M+1}\, \frac{\Gamma (-M,-i r \omega )}{r}+\frac{\,(-1)^M}{M!}\frac{\mathcal{U}(r)}{r}\Bigg]\,,\notag\\[2ex]
&=\sum_{p=1}^{M-1}\bigg(\frac{\Gamma(-M+p-1,-i\,r\,\omega)(-1)^{p}\,(M-p)!}{M!}\,\psi_{M-p+1}\bigg)-\psi_{M+1} \,U(r)+\frac{(-1)^M}{M!}\,\mathcal{V}(r)\,.\notag
\end{align}
The last integral is 
\begin{align}
\mathcal{V}(r)&:=\int^r dr'\,\frac{\mathcal{U}(-i\,r'\,\omega)}{r'}\,,\notag\\[2ex]
&=-i\, r \,\omega  \, _4F_4(1,1,1,1;2,2,2,2;i \,r \,\omega )-\frac{1}{6} \log ^3(-i\, r\, \omega )-\frac{1}{2} \,\gamma_E\,  \log ^2(-i \,r\, \omega)-\frac{1}{2}\,\bigg(\gamma_E^2+\frac{\pi^2}{6}\bigg)\log(-i\,r\,\omega)\,,
\end{align}
where $_4F_4(1,1,1,1;2,2,2,2;z)$ is represented by the convergent series as below,
\begin{align}
_4F_4(1,1,1,1;2,2,2,2;z)=\sum_{k=0}^{\infty}\frac{z^k}{(k+1)^4\,k!}\,.
\end{align}
\subsection{Avoiding Divergences}
In the ealier part of this section, we did not make a particular mention about the boundary condition for the third order contribution $\mathcal{F}_3$ and we even posed the integrations without lower bound, because of divergences. The most problematic term is, 
\begin{align}
U(r)\,r^3\sim r\,\text{e}^{i\,r\,\omega}\,\quad \text{as}\quad r\to\infty\,.
\end{align}
To avoid this problem, we introduce a finite radius $r_d\gg2$. It is supposed that in farther region $r>r_d$, the former approximant $\mathcal{F}_0+\mathcal{F}_1+\mathcal{F}_2$ already provides an excellent approximation so that any additional corrections are redundant. Finally, the solution should look like
\begin{align}
\mathcal{F}=\mathcal{F}_0+\mathcal{F}_1+\mathcal{F}_2+\Theta(r_d-r)\big(\mathcal{F}_3(r)-\mathcal{F}_3(r_d)\big)\,,
\end{align}
where $\Theta(x)$ is a step function such that
$\Theta(x)=1$ if $x>0$, otherwise $\Theta(x)=0$. Or, one could use more smooth function for a bridge instead of $\Theta(x)$. 
\section{Chebyshev interpolation of $\hbar$}\label{sec:hbar}
In this section, we determine the particular value of $\hbar$, which mostly accelerates the convergence of HAM series. In principle, it is permitted to be any complex numbers, however in order to get as accurate solution as possible, we regard $\hbar$ as a variable dependent on $l$. Naturally, it could be a function of both $\omega$ and $l$, but we have found that $\hbar$ is not sensitive to value of $\omega$ empirically. we restrict the regime of $l$ within $l\in[2:10]$ because of a physical interest, and attempt to find a polynomial with Chebyshev nodes \cite{Lloyd2012}. Since $l$ is an integer, only possible simple nodes are $\{2,\,5,\,10\}$ or $\{2,4,8,10\}$, and we choose  $\{2,4,8,10\}$. (Otherwise, we encounter non-integer numbers.)  We determined the value of $\hbar$ by matching our HAM solution and numerical integration, in the cases of $l=\{2,4,8,10\}$ with $\omega=0.01$ at $r=25=\frac{1}{4}\frac{1}{\omega}$ and $r_d=\frac{2}{\omega}$. The reason that the standard values of $\omega$ and $r$ are chosen as $0.01$ and $25$, is that the regime of around $\omega\sim0.01$ is mostly of interest in binaries system and we would like our solution to be valid even when $r\ll\frac{1}{\omega}$. The results are as what follows:
\begin{subequations}
\begin{align}
\hbar=&-(0.0000459883\, -0.000171702 \,i)\, l^3-(0.00784349\, +0.00238014\, i)\, l^2\notag\\[1ex]
&-(0.229561\, +0.0697489\, i)\, l+(-0.952543+0.00250952\, i)\,,\quad(\text{for}\,\,q^1\, \text{order})\,,\\[3ex]
\hbar=&\,(0.000296599\, -0.0000640214\, i)\, l^3-(0.00906544\, -0.00253088\, i)\, l^2\notag\\[1ex]
&-(0.126013\, +0.0837528\, i) \,l+(-0.96776+0.0647684\, i)\,,\quad(\text{for}\,\,q^2\, \text{order})\,,\\[3ex]
\hbar=&\,(0.000353563\, +0.0000221722 \,i) \,l^3-(0.00991384\, -0.000662909 \,i) \,l^2\notag\\[1ex]
&-(0.0705465\, +0.0688801\, i)\, l+(-0.998537+0.0796354\, i)\,,\quad(\text{for}\,\,q^3\, \text{order})\,.
\end{align}
\end{subequations}
We present the comparsion between numerical integrations $\mathcal{F}_n$ and our HAM solutions $\mathcal{F}_h$ for two cases. The vertical axis of Fig.\ref{fig:0.01_2} and Fig.\ref{fig:0.17_5}, indicates $\big|\frac{\mathcal{F}_n-\mathcal{F}_h}{\mathcal{F}_n}\big|$. It is easily seen that as higher order one has, one will get more accurate approximation. Furthermore, as expected, our HAM solution gets less accurate as $\frac{1}{r}$ increases, or closer to the horizon.
\begin{figure}[ht]
\includegraphics[height=0.4\textwidth,width=0.8\textwidth]{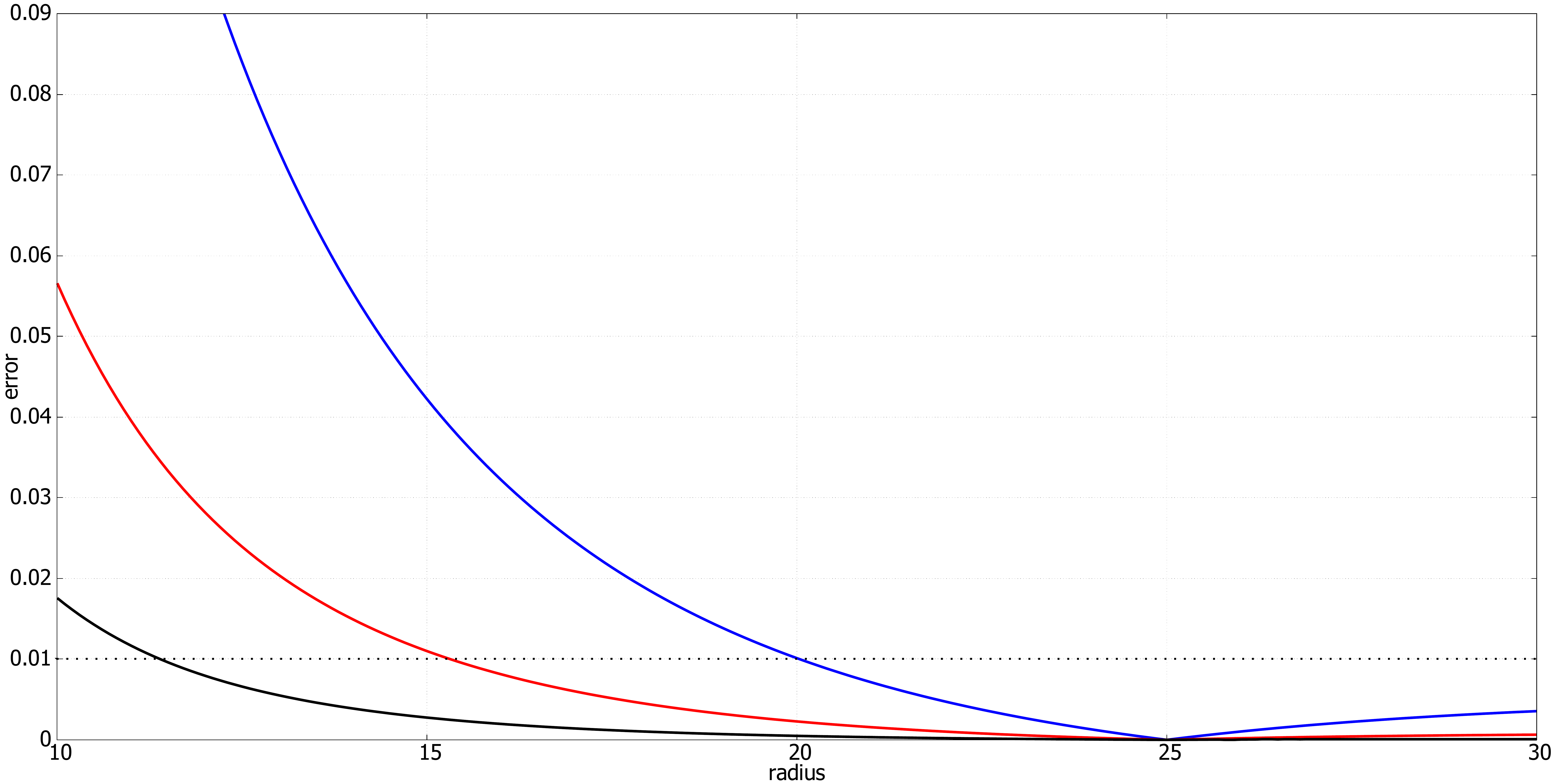}
\centering
\caption{The error $=\big|\frac{\mathcal{F}_n-\mathcal{F}_h}{\mathcal{F}_n}\big|$ is plotted for $\omega=0.01$, $l=2$ case. The {\color{yellow}yellow} line is $\mathcal{F}_0$, the {\color{blue}blue} one is $\mathcal{F}_0+\mathcal{F}_1$. The {\color{red}red} one is $\mathcal{F}_0+\mathcal{F}_1+\mathcal{F}_2$ and the \textbf{black} is $\mathcal{F}_0+\mathcal{F}_1+\mathcal{F}_2+\Theta(r_d-r)\big(\mathcal{F}_3(r)-\mathcal{F}_3(r_d)\big)$. As $\hbar$ constructed, the error at $r=25$ is zero. We presented $\mathcal{F}_0$ but it could not be exbihited because of too large error.}
\label{fig:0.01_2}
\end{figure}
\begin{figure}[ht]
\includegraphics[height=0.4\textwidth,width=0.8\textwidth]{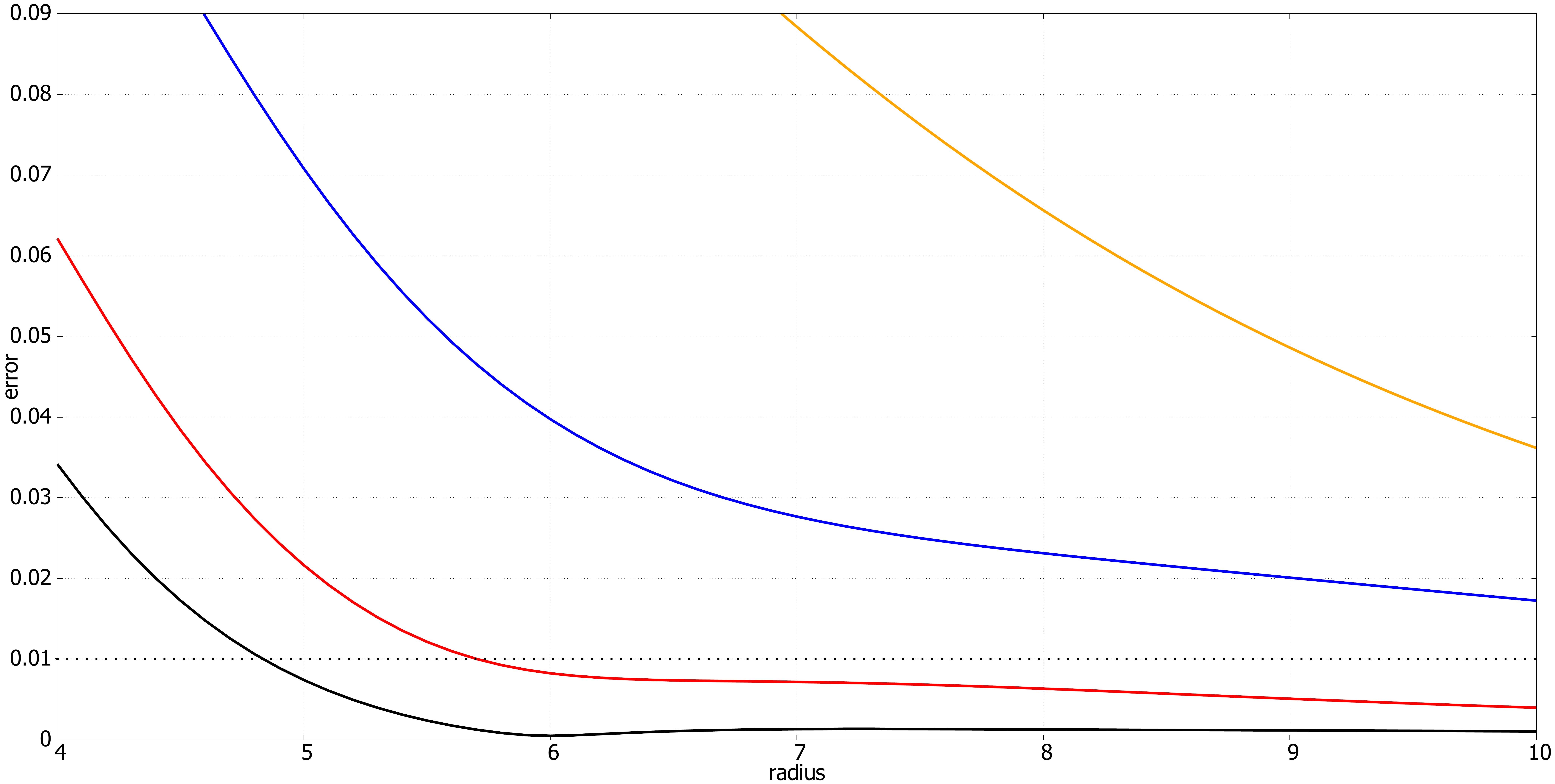}
\centering
\caption{The error $=\big|\frac{\mathcal{F}_n-\mathcal{F}_h}{\mathcal{F}_n}\big|$ is plotted for $\omega=0.01$, $l=2$ case. The {\color{yellow}yellow} line is $\mathcal{F}_0$, the {\color{blue}blue} one is $\mathcal{F}_0+\mathcal{F}_1$. The {\color{red}red} one is $\mathcal{F}_0+\mathcal{F}_1+\mathcal{F}_2$ and the \textbf{black} is $\mathcal{F}_0+\mathcal{F}_1+\mathcal{F}_2+\Theta(r_d-r)\,\Big(\mathcal{F}_3(r)-\mathcal{F}_3(r_d)\Big)$.} 
\label{fig:0.17_5}
\end{figure}
\end{widetext}
\section{Summary and Future Plans}
In this paper, we have derived HAM solution of vacuum Regge-Wheeler equations up to the third order. And the numerical assesments showed that our HAM solution has good agreement with numerical solutions unless it is so near to the horizon. We have seen that the HAM procedure presented here really improves the accuracy of our analytic solution, despite of the possiblity of better way to find more precise numerical choices of $M$,$\hbar$ and $r_d$. Although we empirically suggested how to determine those numbers with decent agreement, we would like to emphsize again that those choices are not necessary.\\
Because our solution is purely analytic, by which we mean that one should not evaluate the values of the functions in the middle points between the infinite boundary and where we are interested in, as well as one is able to be unaware of particular values of variables such as $l$ and $\omega$,  we are able to determine instanteneously what will happen in null-infinity, out of whatever happened within finite regime around blackholes. This will be able to provide us a huge advantage given that every matter sources likely to exist in reality, should be distributed in a spatially finite region, and so the configuration of gravitational field around the sources, must be strongly approximated by a perturbed Schwarzchild or Kerr metric at a certian distance from the sources. This point provokes us to do the very next step that is to extend gravitational field solutions valid only within certian finite region, to infinity. The similiar procedure has been done in the framework of post-Newtonian theory (PN) and post-Minkowskian (PM) formalism, where post-Newtoian metric solution is extended to infinity by a perturbed Minkowskian metric \cite{BlanchetReview,Poujade2002}. We are expecting that replacing the perturbed Minkowskian metric to a perturbed Schwarzchild metric must improve our knowledge on gravitaitonal radiation.\\
\indent Another application is an usage for the self-force theory. In order to  evaluate dynamics of extreme mass-ratio inspiraling binaries (EMRI) sufficiently, it is known that the second-order dissipative self-force and the first-order conservative self-force contribution \cite{Miller2020}. We are also expecting that our HAM solution probably provides an analytic expression for the first-order conserivative orbits around Schwarzchild metric by following the regularizion procedure derived by \cite{Thompson2019} in a gauge condition easy to evaluate, so that one does not need to compute the whole perturbation metric numerically everytime to get how the self-force affects to dynamics of binaries and gravitational radiation. \\
\indent In the perspective of the art of solving differential equations, the feasible future plan should be to (1) extend our HAM solution to higher orders, and (2) to apply HAM to solving Sasaki-Nakamura equation \cite{Sasaki1982} which is a generalizaiton of Regge-Wheeler equation to spinning blackholes. If the latter goes succesful, the solution will enable us to repeat the above two applications (PN/EMRI) in more general perspective.
\section*{Acknowledgement}
Every symbolic/numerical computations here were carried out by the computing program, \textsc{Mathematica}.
\newpage
\begin{appendix}
\begin{widetext}
\section{Reconstructing perturbed Schwarzschild metric from Regge-Wheeler function}\label{app:Rec}
This appendix is designed to present how vacuum outgoing-wave Regge-Wheeler function $Q_{l\omega}(r)$ can reconstruct entire perturbed Schwarzschild metric. Throughout this appendix, it is always supposed that explicit expression of $Q_{l\omega}(r)$ are known. For the first thing, we could build ingoing-wave solution immediately, by $Q_{l-\omega}(r)$ ($\omega\to-\omega$). It is because that Regge-Wheeler equation Eq.(\ref{RWeq}) is invariant under the exchange $\omega\to-\omega$. Thus, we have actually, general solution by linear combinations of $Q_{l\omega}(r)$ and $Q_{l-\omega}(r)$. For another thing, non-homogeneous solution could be built from homogeneous solutions easily in many cases where sources are including Dirac delta function via so called Green's function method. Let $Q_1(r)$ and $Q_2(r)$ be two homognenous solutions with a proper boundary condition at $r\to\infty$ and $r\to2$ respectively. Then, the non-homogeneous solution of 
\begin{align}\label{nonRWeq}
\frac{d^2Q(r)}{dr^{*2}}+\bigg[\,\omega^2-\Big(1-\frac{2}{r}\Big)\,\Big(\frac{l\,(l+1)}{r^2}-\frac{6}{\,r^3}\Big)\bigg]\,Q(r)=S(r)\,,
\end{align}
is 
\begin{align}\label{Green}
Q(r)=\frac{Q_1(r)}{W}\int^{r^*}_{-\infty}\,dr^{*'}\,Q_2(r')\,S(r')+\frac{Q_2(r)}{W}\,\int^\infty_{r^*}\,dr^{*'}\,Q_1(r')\,S(r')\,,
\end{align}
where the Wronskian $W=\frac{dQ_1}{dr^*}\,Q_2-\frac{dQ_2}{dr^*}\,Q_1$. If $S(r)$ has Dirac delta function as in many cases, the integrations are straightforward.\\
\indent Now, let us turn our attention to other gauge invariants such as Zerilli function $Z(r)$ and outgoing radiative Wely scalar $\Psi_4(r)$. The following algebraic relations (generalized Darboux transformation) hold between homogeneous solutions \cite{Chandrasekhar1975,Glampedakis2017},
\begin{subequations}\label{others}
\begin{align}
Z(r)&\propto\Big(1-\frac{2}{r}\Big)\,\frac{dQ}{dr}
+\frac{\lambda ^2\, (\lambda +2)\, r^3+6 \,\lambda \, (\lambda +2) \,r^2+72\, r-144}{12\, r^2 \,(\lambda \, r+6)}\,Q\notag\,,\\[1ex]
\Psi_4(r)&\propto\frac{2\,(r-2)\,(-3+r+i\,r^2\,\omega)}{r^4}\,\frac{dQ}{dr}+\frac{12-2\,r\,(5+\lambda)+r^2\,(2+\lambda-6\,i\,\omega)+2\,i\,r^3\,\omega-2\,r^4\,\omega^2}{r^5}\,Q\,.\notag
\end{align}
\end{subequations}
Note that $\lambda=(l-1)\,(l+2)$ and, `$\propto$' means `proportional to' as implying that the above relations cannot determine their overall amplitudes. From these homogeneous solutions, it is always possible to get their non-homogeneous counterparts in the same way as Eq.(\ref{Green}).\\
Now we construct the entire metric $g_{\mu\nu}=g^{0}_{\mu\nu}+h_{\mu\nu}$, where $g^{0}_{\mu\nu}$ is Schwarzschild metric in Schwarzschild coordinate. Once $Q(r)$ and $Z(r)$ are determined, the 6 gauge-invariants $\{\alpha,\,\beta,\,\chi,\,\phi,\,\delta,\,\epsilon\}$ are determined as defined in Eqs.(\ref{invariants}), and once the invariants are determined, every components of metric are determined as long as a gauge fixing condition is given. Here, for convenience of readers, we list the results of \cite{Thompson2017}:
\begin{subequations}\label{invariants}
\begin{align}
\alpha&=-\frac{r}{\lambda}\,\bigg[\frac{-i\,\omega}{r-2}\,Q+r\,E_J\bigg]\,,\\[2ex]
\beta&=-\frac{1}{\lambda}\bigg[\frac{r-2}{r}\partial_rQ-\frac{r-2}{\lambda\,r^2}Q-r^2\,E_C\bigg]\,,\\[2ex]
\chi&=\frac{-1}{(\lambda+2)(\lambda\,r+6)}\bigg[2(\lambda\,r+6)\partial_rZ+\frac{\lambda(\lambda+2)r^2+6(r\,\lambda+4)}{r\,(r-2)}Z+\frac{r^4}{r-2}\,E_A\bigg]\,,\\[2ex]
\psi&=\frac{-1}{(r-2)^2(\lambda \,r+6)}\bigg[r(r-2)(\lambda r+6)\partial_rZ+(r^2\lambda-3r\lambda-6)\,Z+\frac{r}{2}\,E_A\bigg]\,,\\[2ex]
\delta&=\frac{r^2}{\lambda+2}\,\bigg[-E_D+\frac{4\,(r-2)}{r^2}(-i\,\omega)\psi-\frac{\lambda+2}{r}\,(-i\,\omega)\,\chi\bigg]\,,\\[2ex]
\epsilon&=\frac{r\,(r-2)}{2}\,\bigg[E_F-\frac{2(r-2)}{r^3}\psi+\frac{2}{r^2}\,\chi+\frac{2\,(r-2)}{r^2}\,\partial_r\chi\bigg]\,.
\end{align}
\end{subequations}
And also the 6 gauge-invaraints are expressed in terms of harmonic modes of metric components as
\begin{subequations}
\begin{align}
\alpha&=J-\frac{r}{2}\partial_rG\,,\\[1ex]
\beta&=-C-\frac{r}{2}\partial_tG\,,\\[1ex]
\chi&=H-\frac{r}{2(r-2\,m)}E-\frac{l(l+1)\,r}{4\,(r-2\,m)}F-\frac{r}{2}\,\partial_rF\,,\\[1ex]
\psi&=\frac{1}{2}\,K-\frac{r\,(r-3)}{2\,(r-2\,m)}E-\frac{r^2}{2\,(r-2\,m)}\partial_rE-\frac{l\,(l+1)\,r\,(r-3\,m)}{4\,(r-2\,m)^2}F-\frac{l\,(l+1)\,r^2}{4\,(r-2)}\partial_rF\,,\\[1ex]
\delta&=D+\frac{r^2}{2\,(r-2)}\partial_tE-\frac{r-4\,m}{r-2\,m}B-r\,\partial_rB-\frac{r^2}{2}\,\frac{\partial^2}{\partial_t\partial_r}F+r\,\frac{4(3\,m-r)+r\,l\,(l+1)}{4\,(r-2\,m)}\partial_tF\,,\\[1ex]
\epsilon&=-\frac{1}{2}A-\frac{m}{2\,r}E-r\partial_tB-\frac{l(l+1)m}{4\,r}F-\frac{r^2}{2}\partial_t^2F\,.
\end{align}
\end{subequations}
Note that by Eqs.(\ref{invariants}), the ten harmonic modes $\{A,B,C,D,E,F,G,H,J,K\}$ linearly determine six variables $\{\alpha,\,\beta,\,\chi,\,\phi,\,\delta,\,\epsilon\}$ but not vice versa. (One needs 4 four more equations \textit{i.e.} gauge conditions.) Finally, the harmonic modes compose the metric components $h_{\mu\nu}$ with the spherical harmonic function $Y^{lm}(\theta,\phi)$ as
\begin{subequations}
\begin{align}
h_{tt}&=A\,Y^{lm}\,,\\[1ex]
h_{tr}&=-D\,Y^{lm}\,,\\[1ex]
h_{rr}&=K\,Y^{lm}\,,\\[1ex]
h_{t\theta}&=-B\,(r\,\partial_\theta Y^{lm})+C\,\left(\frac{r}{\sin\theta}\partial_\phi Y^{lm}\right)\,,\\[1ex]
h_{t\phi}&=-C\,(r\,\sin\theta\,\partial_\theta Y^{lm})-B\,\left(r\,\partial_\phi Y^{lm}\right)\,,\\[1ex]
h_{r\theta}&=H\,(r\,\partial_\theta Y^{lm})-J\,\left(\frac{r}{\sin\theta}\partial_\phi Y^{lm}\right)\,,\\[1ex]
h_{r\phi}&=J\,(r\,\sin\theta\,\partial_\theta Y^{lm})+H\,\left(r\,\partial_\phi Y^{lm}\right)\,,\\[1ex]
h_{\theta\theta}&=\Big(E+F\,\big(\partial_\theta^2+\frac{1}{2}l(l+1)\big)\Big)(r^2\,Y^{lm})-G\,\frac{r^2}{\sin\theta}\,\big(\partial_\theta\partial_\phi-\cot\theta\,\partial_\phi\big)\,Y^{lm}\,,\\[1ex]
h_{\theta\phi}&=r^2\,F\Big(\partial_\theta\partial_\phi-\cot\theta\,\partial_\phi\Big)\,Y^{lm}-\frac{r^2}{2}\,G\,\Big(\frac{1}{\sin\theta}\partial_\phi^2+\cos^2\theta\,\partial_\theta-\sin\,\partial_\theta^2\Big)\,Y^{lm}\,,\\[1ex]
h_{\phi\phi}&=\sin^2\theta\,\Big(E-F\,\big(\partial_\theta^2+\frac{1}{2}l(l+1)\big)\Big)\,(r^2\,Y^{lm})-r^2\,G\,\big(\partial_\theta\partial_\phi-\cot\theta\,\partial_\phi\big)\,Y^{lm}\,.
\end{align}
\end{subequations}
Similiarly,
\begin{subequations}
\begin{align}
-16\,\pi\,T_{tt}&=E_A\,Y^{lm}\,,\\[1ex]
-16\,\pi\,T_{tr}&=-E_D\,Y^{lm}\,,\\[1ex]
-16\,\pi\,T_{rr}&=E_K\,Y^{lm}\,,\\[1ex]
-16\,\pi\,T_{t\theta}&=-E_B\,(r\,\partial_\theta Y^{lm})+E_C\,\left(\frac{r}{\sin\theta}\partial_\phi Y^{lm}\right)\,,\\[1ex]
-16\,\pi\,T_{t\phi}&=-E_C\,(r\,\sin\theta\,\partial_\theta Y^{lm})-E_B\,\left(r\,\partial_\phi Y^{lm}\right)\,,\\[1ex]
-16\,\pi\,T_{r\theta}&=E_H\,(r\,\partial_\theta Y^{lm})-E_J\,\left(\frac{r}{\sin\theta}\partial_\phi Y^{lm}\right)\,,\\[1ex]
-16\,\pi\,T_{r\phi}&=E_J\,(r\,\sin\theta\,\partial_\theta Y^{lm})+E_H\,\left(r\,\partial_\phi Y^{lm}\right)\,,\\[1ex]
-16\,\pi\,T_{\theta\theta}&=\Big(E_E+E_F\,\big(\partial_\theta^2+\frac{1}{2}l(l+1)\big)\Big)(r^2\,Y^{lm})-E_G\,\frac{r^2}{\sin\theta}\,\big(\partial_\theta\partial_\phi-\cot\theta\,\partial_\phi\big)\,Y^{lm}\,,\\[1ex]
-16\,\pi\,T_{\theta\phi}&=r^2\,E_F\Big(\partial_\theta\partial_\phi-\cot\theta\,\partial_\phi\Big)\,Y^{lm}-\frac{r^2}{2}\,E_G\,\Big(\frac{1}{\sin\theta}\partial_\phi^2+\cos^2\theta\,\partial_\theta-\sin\,\partial_\theta^2\Big)\,Y^{lm}\,,\\[1ex]
-16\,\pi\,T_{\phi\phi}&=\sin^2\theta\,\Big(E_E-E_F\,\big(\partial_\theta^2+\frac{1}{2}l(l+1)\big)\Big)\,(r^2\,Y^{lm})-r^2\,E_G\,\big(\partial_\theta\partial_\phi-\cot\theta\,\partial_\phi\big)\,Y^{lm}\,.
\end{align}
\end{subequations}
\section{Explicit expressions of the coefficients}\label{app:Coe}
The coefficients in $\mathcal{G}_1$ are
\begin{subequations}
\begin{align}
\gamma_0&=-\frac{\,\hbar\, \omega ^2 (\,\mathcal{A}\, (M+2)+i \,\mathcal{B}\, \omega )}{(M+1) (M+2)}\,,\\[1ex]
\gamma_1&=\frac{\,\hbar\, \omega ^2 (-2 \,\mathcal{A}\, (M+2)-2 i \,\mathcal{B}\, \omega +\,\mathcal{B}\,)}{(M+1) (M+2)}\,,\\[1ex]
\gamma_2&=\frac{\,\hbar\, (\,\mathcal{A}\, (M+2) (M+P-4 i \omega +1)+\,\mathcal{B}\, \omega  (6 \omega +i P))}{(M+1) (M+2)}\,,\\[1ex]
\gamma_3&=-\frac{\,\hbar\, \left(2 \,\mathcal{A}\, \left(M^2+7 M+10\right)+\,\mathcal{B}\, \left(-M^2+M (-3+4 i \omega )+P+12 i \omega -2\right)\right)}{(M+1) \,(M+2)}\,,\\[1ex]
\gamma_4&=-\frac{2 \,\mathcal{B}\, \,\hbar\, M (M+4)}{(M+1)\, (M+2)}\,,\\[1ex]
\delta_0&=\frac{\,\hbar\, (-i)^M \omega ^{M+3} (\,\mathcal{B}\, \omega -i \,\mathcal{A}\, (M+2))}{(M+1) (M+2)}\,,\\[1ex]
\delta_1&=\frac{\,\hbar\, (-i)^M \omega ^{M+2} (\,\mathcal{A}\, (M+2) (M-2 i \omega +1)+\,\mathcal{B}\, \omega  (i M+2 \omega +2 i))}{(M+1) (M+2)}\,,\\[1ex]
\delta_2&=\frac{\,\hbar\, (-i)^M \omega ^{M+1} (\,\mathcal{A}\, (M+2) (2 (M+3) \omega +i P)+i \,\mathcal{B}\, \omega  (2 (M+4) \omega +i P))}{(M+1) (M+2)}\,,\\[1ex]
\delta_3&=-\frac{\,\hbar\, (-i)^M \omega ^M (\,\mathcal{A}\, (M+2) (M P+P+8 i \omega )+i \,\mathcal{B}\, \omega  ((M+2) P+8 i \omega ))}{(M+1) (M+2)}\,,\\[1ex]
\delta_4&=\frac{6 \,\hbar\, (-i)^M \omega ^M (\,\mathcal{A}\, M+\,\mathcal{A}\,+i \,\mathcal{B}\, \omega )}{M+1}\,.
\end{align}
\end{subequations}
\\
The explicit expressions of the coefficients in $\mathcal{H}_2$ are
\begin{subequations}
\begin{align}
\epsilon_0&=-\frac{\,\hbar\, \omega ^2 (\,\mathcal{A}\, (M+2)+i \,\mathcal{B}\, \omega )}{2 (M+1) (M+2)}\,,\\[1ex]
\epsilon_1&=\frac{\,\hbar\, \omega  (\,\mathcal{B}\, \omega  (M-4 i \omega +3)-i \,\mathcal{A}\, (M+2) (M-4 i \omega +1))}{2 (M+1) (M+2)}\,,\\[1ex]
\epsilon_2&=\frac{\,\hbar\, (\,\mathcal{B}\, \omega  (-(M (7 M+37)+42) \omega -i (M+3) (M ((M+5) P+M+3)+5 P+2))}{(M+1)^2 (M+2)^2 (M+3)}\\[0.5ex]
&\,-\,\hbar\, \,\mathcal{A}\,\frac{(M+2) (M+3) ((M+2) ((M+2) P+M+1)-i (M+5) \omega ))}{(M+1)^2 (M+2)^2 (M+3)}\,,\\[1ex]
\epsilon_3&=\frac{\,\mathcal{A}\, \,\hbar\, (5 M+13)}{(M+1) (M+2)}\\[0.5ex]
&\,+\frac{\,\mathcal{B}\, \,\hbar\, ((M+3) P+i (M (9 M+47)+54) \omega -(M+1) (M+2) (M+3))}{(M+1) (M+2)^2 (M+3)}\notag\,,\\[1ex]
\epsilon_4&=\frac{2 \,\mathcal{B}\, \,\hbar\, M (M+4)}{(M+1)\, (M+2) \,(M+3)}\,,\\[1ex]
\zeta_0&=\frac{\,\hbar\, (-i)^M \omega ^{M+3} (\,\mathcal{B}\, \omega -i \,\mathcal{A}\, (M+2))}{2 (M+1) (M+2)}\,,\\[1ex]
\zeta_1&=\frac{\,\hbar\, (-i)^M \omega ^{M+2} (\,\mathcal{A}\, (M+2) (M-2 i \omega +1)+\,\mathcal{B}\, \omega  (2 \omega +i (M+2)))}{(M+1) (M+2)}\,,\\[1ex]
\zeta_2&=\frac{i \,\mathcal{A}\, \,\hbar\, (-i)^M \omega ^{M+1} \left(M^4+M^3 (4-4 i \omega )+M^2 (-2 P-16 i \omega +3)-2 M (4 P+9 i \omega +2)-8 P+2 i \omega -4\right)}{2 (M+1)^2 (M+2)}\notag\\[0.5ex]
&\,-\frac{\,\mathcal{B}\, \,\hbar\, (-i)^M (M (M+1) (M+2) (M+3)-2 (M (M+5)+5) P) \omega ^{M+2}}{2 (M+1)^2 (M+2)^2}\notag\\[0.5ex]
&\,+\frac{i \,\mathcal{B}\, \,\hbar\, (-i)^M (M (M (2 M (M+7)+27)-3)-30) \omega ^{M+3}}{(M+1)^2 (M+2)^2 (M+3)}\,,\\[1ex]
\zeta_3&=\frac{\,\hbar\, (-i)^M \omega ^M \left(i (M+2) \omega  (8 \,\mathcal{A}\,+\,\mathcal{B}\, P)+\,\mathcal{A}\, (M+1) (M+2) P-8 \,\mathcal{B}\, \omega ^2\right)}{(M+1) (M+2)}\,,\\[1ex]
\zeta_4&=-\frac{3 \,\hbar\, (-i)^M \omega ^M (\,\mathcal{A}\, M+\,\mathcal{A}\,+i \,\mathcal{B}\, \omega )}{M+1}\,.
\end{align}
\end{subequations}
\\

The explicit expressions of the coefficients in $\mathcal{I}_2$ are
\begin{subequations}
\begin{align}
\eta_0&=-\frac{\,\hbar\, \omega ^2 (\,\mathcal{A}\, (M+2)+i \,\mathcal{B}\, \omega )}{3 (M+1) (M+2)}\,,\\[1ex]
\eta_1&=\frac{\,\hbar\, \omega  (\,\mathcal{B}\, \omega  (M-6 i \omega +4)-i \,\mathcal{A}\, (M+2) (M-6 i \omega +1))}{6 (M+1) (M+2)}\,,\\[1ex]
\eta_2&=-\frac{\,\hbar\, \left(\,\mathcal{A}\, (M+2) \left(M^2+6 i M \omega -6 P+30 i \omega -1\right)+i \,\mathcal{B}\, \omega  \left(M^2+M (3+6 i \omega )-6 P+42 i \omega +2\right)\right)}{6 (M+1) (M+2)}\,,\\[1ex]
\eta_3&=\frac{8 \,\mathcal{A}\, \,\hbar\, (M+2)^3+\,\mathcal{B}\, \,\hbar\, ((M+2) (P-(M+1) (M+2))+4 i (M+3) (3 M+4) \omega )}{\left(M^2+3 M+2\right)^2}\,,\\[1ex]
\eta_4&=\frac{2 \,\mathcal{B}\, \,\hbar\, M (M+4)}{(M+1) (M+2)^2}\,,\\[1ex]
\theta_0&=\frac{\,\hbar\, (-i)^M \omega ^{M+3} (\,\mathcal{B}\, \omega -i \,\mathcal{A}\, (M+2))}{3 (M+1) (M+2)}\,,\\[1ex]
\theta_1&=\frac{\,\hbar\, (-i)^M \omega ^{M+2} (\,\mathcal{A}\, (M+2) (M-2 i \omega +1)+\,\mathcal{B}\, \omega  (i M+2 \omega +2 i))}{2 (M+1) (M+2)}\,,\\[1ex]
\theta_2&=\frac{\,\hbar\, (-i)^M \omega ^{M+1} (\,\mathcal{A}\, (M+2) (2 (M+3) \omega +i P)+i \,\mathcal{B}\, \omega  (2 (M+4) \omega +i P))}{(M+1) (M+2)}\,,\\[1ex]
\theta_3&=\frac{\,\mathcal{A}\, \,\hbar\, (-i)^M \omega ^M \left((M+1)^2 ((M-1) M-6 (P+1))+6 i (M (M (M+6)+17)+20) \omega \right)}{6 (M+1)^2}\\[0.5ex]
&\,+\frac{i \,\mathcal{B}\, \,\hbar\, (-i)^M \omega ^{M+1} \left((M+2)^2 (M (M (M+2)-6 P-5)-6)+6 i (M (M (M (M+10)+41)+84)+60) \omega \right)}{6 (M+1)^2 (M+2)^2}\notag\,,\\[1ex]
\theta_4&=-\frac{6 \,\hbar\, (-i)^M \omega ^M (\,\mathcal{A}\, M+\,\mathcal{A}\,+i \,\mathcal{B}\, \omega )}{M+1}\,.
\end{align}
\end{subequations}
Note that $P=l\,(l+1)+2\,i\,\omega$.
\section{Asymptotic behaviour of $\mathcal{U}$}\label{app:AsympU}
This section provides asymptotic value of $\mathcal{U}(z)$. It is begun by presenting the integral representation \cite{HPQintegral} of $_3F_3(1,1,1;2,2,2,z)$,
\begin{align}
_3F_3(1,1,1;2,2,2,z)=\int^{1}_0 dt\, _2F_2(1,1;2,2,z\,t)\,,
\end{align}
where $_2F_2(1,1;2,2,z\,t)=-\frac{\gamma_E+\Gamma(0,-z\,t )+\log (-z\,t )}{z\,t}$. Hence,
\begin{align}
\lim_{r\to+\infty}\,_3F_3(1,1,1;2,2,2,i\,r)&=\lim_{r\to+\infty}\int^{1}_0 \frac{\gamma_E+\Gamma(0,-i\,r\,t)+\log(-i\,r\,t)}{-i\,r\,t}  dt\,.
\end{align}
Note that $\omega$ is omitted here, which was supposed to be included via $z=-i\,r\,\omega$, because it does not affect to asymptotic behaviour if $\omega>0$.
We introduced a small positive parameter $\epsilon$ for the convenience of handling divergent terms. Hereafter, $\lim_{r\to+\infty}$, $\lim_{\epsilon\to+\infty}$ are occasionally omitted unless there is confusing. By a change of variable $\kappa:=r\,t$, the integration becomes
\begin{align}\label{asym3F3}
\lim_{r\to+\infty}\,_3F_3(1,1,1;2,2,2,i\,r)&=\lim_{r\to+\infty}\lim_{\epsilon\to0+}\frac{i}{r}\int^{r}_\epsilon  d\kappa\frac{\gamma_E+\Gamma(0,-i\,\kappa)+\log(-i\,\kappa)}{\,\kappa } \,,\\[1ex]\notag
&=\lim_{r\to+\infty}\frac{i}{r}\lim_{\epsilon\to0+}\Bigg[\gamma_E\,\log(r/\epsilon)+\int^{r}_\epsilon d\kappa\frac{\Gamma(0,-i\,\kappa)}{\kappa}+\frac{1}{2}(\log^2(-i\, r)-\log^2(-i\,\epsilon))\Bigg]\,.
\end{align}
Let us concentrate on the second term $\int^{r}_\epsilon d\kappa\frac{\Gamma(0,-i\,\kappa)}{\kappa}$. By the Cauchy integral formula, it could be found that the contour integration vanishes
\begin{align}\label{contour}
\oint_\mathcal{C} d\kappa\frac{\Gamma(0,-i\,\kappa)}{\kappa}=0\,,
\end{align}
where the contour $\mathcal{C}$ is presented in Fig.\ref{fig:contour}.
\begin{figure}[t]
\includegraphics[height=0.35\textwidth,width=0.5\textwidth]{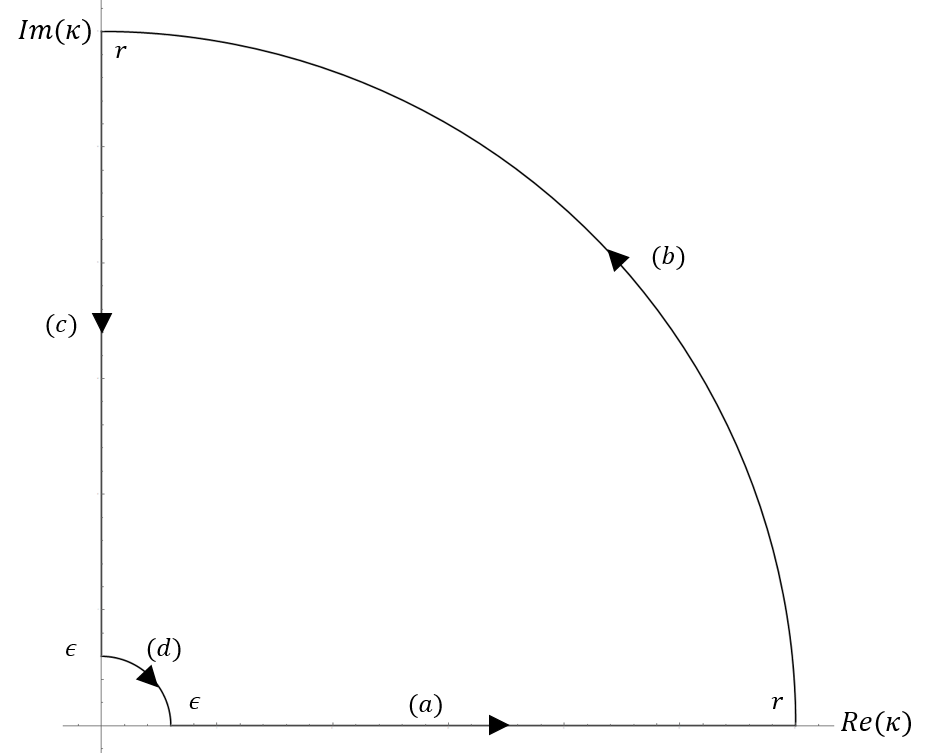}
\centering
\caption{The contour $\mathcal{C}$ is composed of $(a)$, $(b)$, $(c)$ and $(d)$ clockwise.}
\label{fig:contour}
\end{figure}
The integration Eq.\ref{contour} could be splitted into the four pieces,
\begin{align}
0=\underbrace{\int^{r}_\epsilon d\kappa\frac{\Gamma(0,-i\,\kappa)}{\kappa}}_{(a)}+\underbrace{i\int^{\pi/2}_{0}d\theta\,\Gamma(0,-i \,r\,\text{e}^{i\,\theta})}_{(b)}+\underbrace{\int_{i r}^{i\epsilon} d\kappa\frac{\Gamma(0,-i\,\kappa)}{\kappa}}_{(c)}+\underbrace{i\int_{\pi/2}^{0}d\theta\,\Gamma(0,-i \,\epsilon\,\text{e}^{i\,\theta})}_{(d)}\,.
\end{align}
From Eq.(\ref{asymGamma}), the integrand of $(b)$ is
\begin{align}
\lim_{r\to\infty}\Gamma(0,-i\,r\,\text{e}^{i\,\theta})=\lim_{r\to\infty}\frac{\text{e}^{i\,r\,\cos\theta-r\,\sin\theta}}{-i\,r\,\text{e}^{i\,\theta}}=0,
\end{align}
and which, hence, leads to $(b)=0$. And in the case of $(d)$, the part of the integrand not vanishing is
\begin{align}
\lim_{\epsilon\to0+}\Gamma(0,-i\,\epsilon\,\text{e}^{i\,\theta})=\lim_{\epsilon\to0+}\big(-\log\epsilon-i\,\theta-\gamma +\frac{i \pi }{2}\big)\,,
\end{align}
which also leads to that $(d)=\frac{1}{8} \pi \big(4 i \log \epsilon +4 i \gamma_E +\pi \big)$. Finally, let us consider the case of $(c)$, 
\begin{align}
\lim_{r\to\infty}(c)&=-\int^\infty_\epsilon\int^\infty_t\,\frac{\text{e}^{-k}}{t\,k}\,dk\,dt=-\int^\infty_\epsilon\int^k_\epsilon\,\frac{\text{e}^{-k}}{t\,k}\,dt\,dk
=-\int^\infty_\epsilon\,\frac{\text{e}^{-k}}{k}\log\frac{k}{\epsilon}\,dk\,,\\[1ex]
&=-\underbrace{\Big[\text{e}^{-k}\log k\,\log\frac{k}{\epsilon}\Big]^{\infty}_\epsilon}_{=0}+\underbrace{\int^\infty_\epsilon\,\frac{\text{e}^{-k}}{k}\log\frac{k}{\epsilon}\,dk}_{=-\lim_{r\to\infty}(c)}+\int^\infty_\epsilon\,\frac{\text{e}^{-k}}{k}\log\epsilon\,dk-\int^{\infty}_\epsilon\,\text{e}^{-k} \log k \log \left(\frac{k}{\epsilon}\right)\notag\,,\notag\\[1ex]
&=\frac{1}{2}\,\int^\infty_\epsilon\,\frac{\text{e}^{-k}}{k}\log\epsilon\,dk-\frac{1}{2}\int^{\infty}_\epsilon\,\text{e}^{-k} \log k \log \left(\frac{k}{\epsilon}\right)\,,\notag\\[1ex]
&=-\,\frac{1}{2}\,\lim_{\epsilon\to0+}\,\bigg(\log^2\epsilon+2\,\gamma_E\,\log\epsilon+\Big(\gamma_E^2+\frac{\,\pi^2}{6}\Big)\bigg)\notag\,.
\end{align}
Therefore, since $(a)=-(b)-(c)-(d)$, we arrive at
\begin{align}
\lim_{r\to\infty}\lim_{\epsilon\to0+}\int^{r}_\epsilon d\kappa\frac{\Gamma(0,-i\,\kappa)}{\kappa}=\frac{1}{2}\bigg(\log^2\epsilon+2\,\gamma_E\,\log\epsilon+\Big(\gamma_E^2+\frac{\,\pi^2}{6}\Big)\bigg)-\frac{1}{8} \pi \big(4 i \log \epsilon +4 i \gamma_E +\pi \big)\,,
\end{align}
which is divergent as $\epsilon\to0$. Now let us go back to Eq.(\ref{asym3F3}). The inside of the sqaure bracket of Eq.(\ref{asym3F3}), becomes 
\begin{align}
&\,\gamma_E\,\log r+\frac{1}{2}\log^2\epsilon+\frac{1}{2}\Big(\gamma_E^2+\frac{\,\pi^2}{6}\Big)-\frac{i\,\pi\,\log\epsilon}{2}-\frac{i\,\gamma_E\,\pi}{2}-\frac{\pi^2}{8}+\frac{1}{2}\log^2(-i\,r)-\frac{1}{2} \log ^2\epsilon+\frac{1}{2} i \pi  \log\epsilon+\frac{\pi ^2}{8}\,,\notag\\[1ex]
=&\,\gamma_E\,\log (-i\,r)+\frac{1}{2}\Big(\gamma_E^2+\frac{\,\pi^2}{6}\Big)+\frac{1}{2}\log^2(-i\,r)\,.
\end{align}
After moving the divergent logarithm terms to the left hand side, one gets
\begin{align}
\lim_{r\to+\infty}\Big[-i\,r\,_3F_3(1,1,1;2,2,2,i\,r)-\gamma_E\,\log (-i\,r)-\frac{1}{2}\log^2(-i\,r)\Big]=\frac{1}{2}\,\Big(\gamma_E^2+\frac{\,\pi^2}{6}\Big)\,,
\end{align}
where the inside of the sqaure bracket of the left hand side is nothing but hte functional part of $\mathcal{U}(\frac{r}{\omega})$ (when $\omega>0$). The simliar calulation reveals that 
\begin{align}
\lim_{r\to-\infty}\Big[-i\,r\,_3F_3(1,1,1;2,2,2,i\,r)-\gamma_E\,\log (-i\,r)-\frac{1}{2}\log^2(-i\,r)\Big]=\frac{1}{2}\,\Big(\gamma_E^2+\frac{\,\pi^2}{6}\Big)\,.
\end{align}
Hence, we can finally conclude that whatever the sign of $\omega\neq0$, the follow holds,
\begin{align}
\lim_{r\to\infty}\mathcal{U}(r)=0\,.
\end{align}
\end{widetext}
\end{appendix}
\bibliographystyle{apsrev}
\bibliography{references.bib}
\end{document}